\documentclass[12pt]{article}
\usepackage{graphicx,amsmath,amssymb,cite}

\newlength{\dinwidth}
\newlength{\dinmargin}
\newcommand{\xB}{x_{\scriptscriptstyle{\mathrm{Bj}}}}
\newcommand{\Pom}{{\mathbb{P}}}
\newcommand{\Reg}{\mathbb{R}}
\newcommand{\dif}{\mathrm{d}}

\begin{document}
\titlepage
\begin{flushright}
  IPPP/06/66          \\
  DCPT/06/132         \\
  7th November 2006   \\
\end{flushright}

\vspace*{0.5cm}

\begin{center}

  {\Large \bf Diffractive parton distributions from H1 data}

  \vspace*{1cm}

  \textsc{A.D. Martin$^a$, M.G. Ryskin$^{a,b}$ and G. Watt$^c$} \\

  \vspace*{0.5cm}
  $^a$ Institute for Particle Physics Phenomenology, University of Durham, DH1 3LE, UK \\
  $^b$ Petersburg Nuclear Physics Institute, Gatchina, St.~Petersburg, 188300, Russia \\
  $^c$ Department of Physics \& Astronomy, University College London, WC1E 6BT,
UK

\end{center}

\vspace*{0.5cm}

\begin{abstract}
  We analyse the latest H1 large rapidity gap data to obtain diffractive parton distributions, using a procedure based on perturbative QCD, and compare them with distributions obtained from the simplified Regge factorisation type of analysis.  The diffractive parton densities and structure functions are made publically available.
\end{abstract}

The H1 Collaboration have recently released measurements of diffractive deep-inelastic scattering (DDIS), $\gamma^* p\to X + p$, with more precision than hitherto.  They have presented data collected using their forward proton spectrometer (FPS) \cite{H1FPS} and also data selected using the large rapidity gap (LRG) method \cite{H1LRG}.  In addition they have used their LRG data, which has much more statistics, to determine diffractive parton densities assuming that Regge factorisation\footnote{The H1 Collaboration call this ``proton vertex factorisation'' \cite{H1LRG}.} holds.  We call this a ``Regge'' fit\footnote{This is the conventional approach based on \emph{collinear factorisation} for the scale dependence \cite{Collins:1997sr} and \emph{Regge factorisation} for the $x_\Pom$ dependence \cite{Ingelman:1984ns}.}.  In principle, this Regge factorisation is \emph{not} justified theoretically, and the purpose of this note is to present diffractive parton distributions obtained from exactly the {\it same} H1 LRG data \cite{H1LRG} using a more theoretically sound procedure.  It is convenient to call this the perturbative QCD (``pQCD'') procedure, although it also contains non-perturbative Regge-like contributions.

\begin{figure}
  \centering
  (a)\hspace{0.3\textwidth}(b)\hspace{0.3\textwidth}(c)\\
  \begin{minipage}{0.3\textwidth}
    \includegraphics[width=\textwidth]{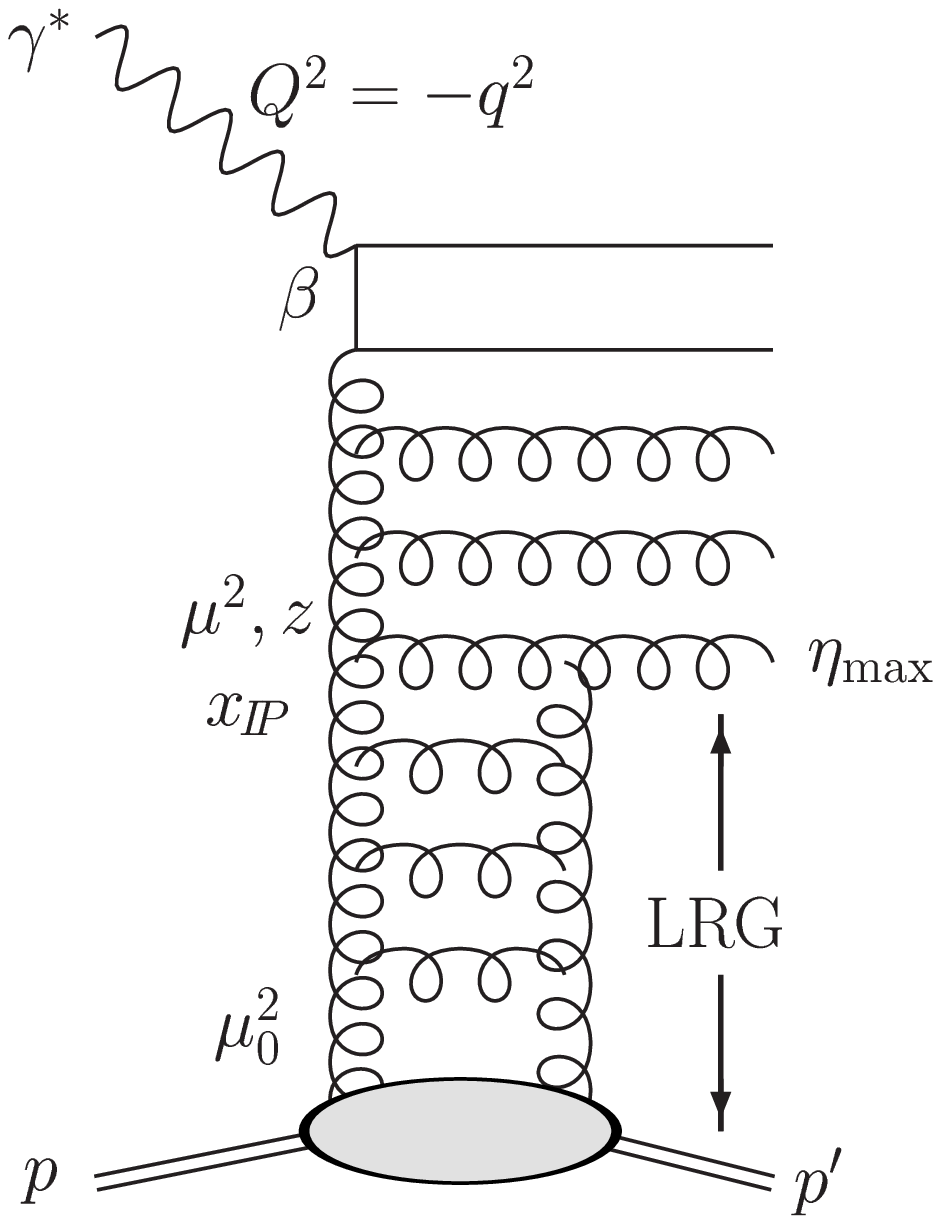}
  \end{minipage}\hfill
  \begin{minipage}{0.3\textwidth}
    \includegraphics[width=0.8\textwidth]{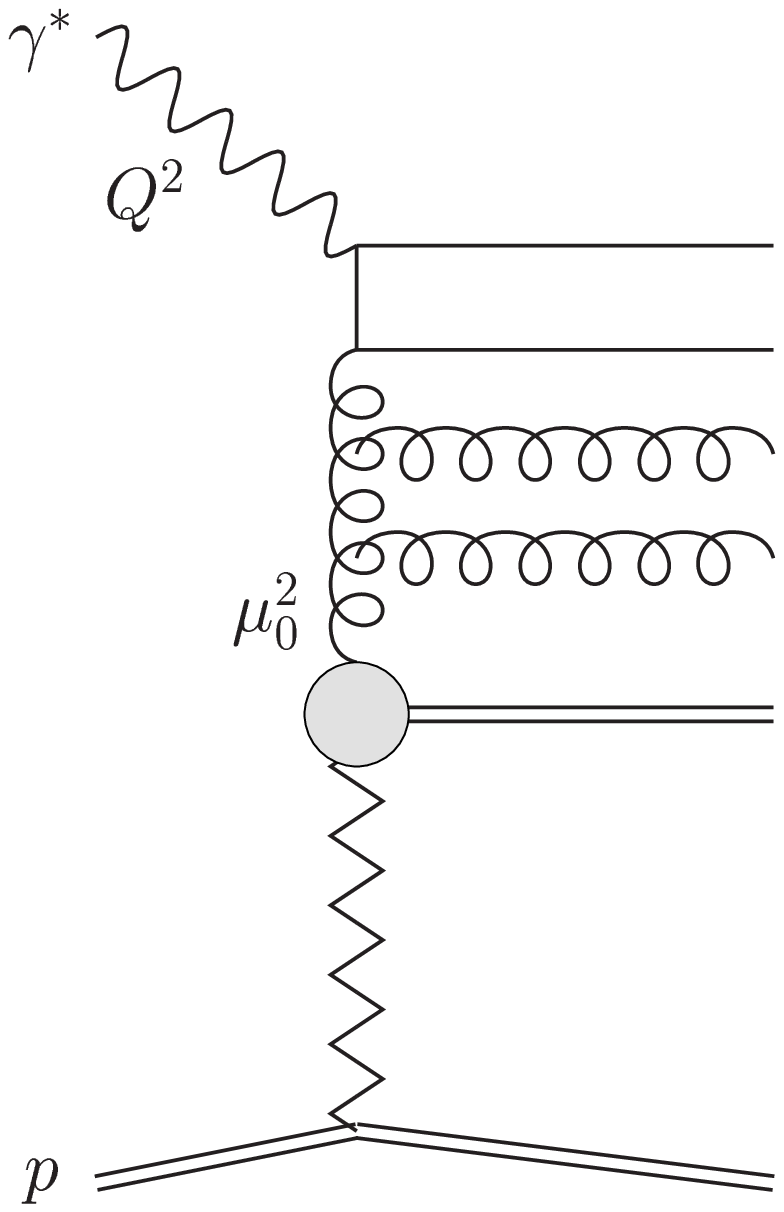}
  \end{minipage}\hfill
  \begin{minipage}{0.3\textwidth}
    \includegraphics[width=\textwidth]{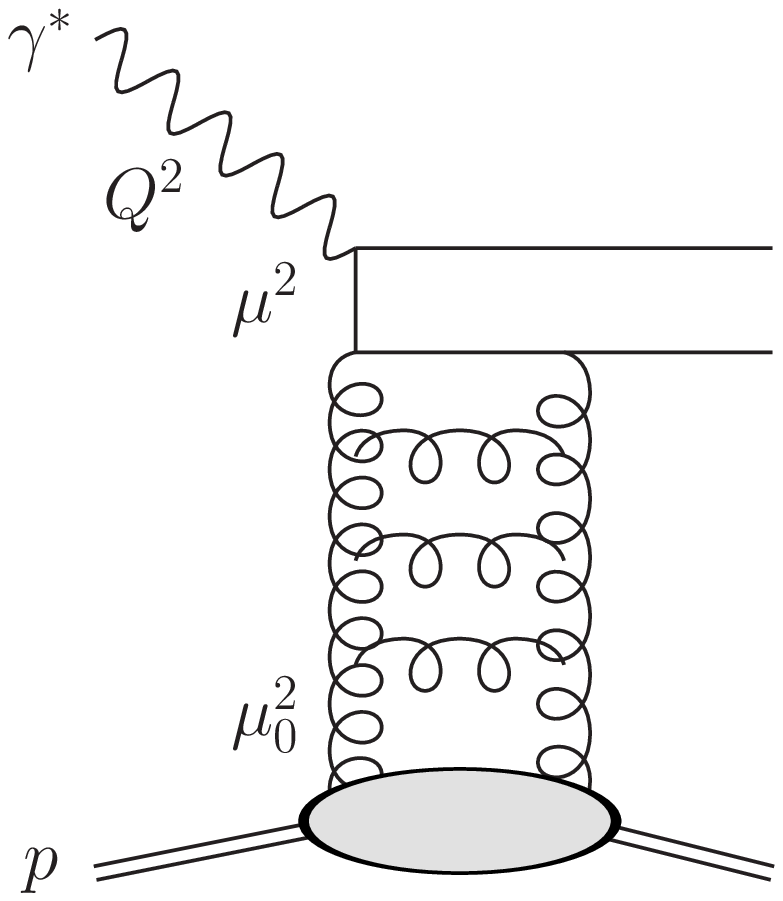}
  \end{minipage}
  \caption{(a) The perturbative \emph{resolved} Pomeron contribution, which is the basis of the pQCD approach,  (b) the non-perturbative \emph{resolved} Pomeron diagram, which accounts for the contribution from low scales, $\mu^2<\mu_0^2$, and (c) the perturbative \emph{direct} Pomeron contribution.}
  \label{fig:diagrams}
\end{figure}

We present the pQCD procedure first.  It is shown schematically in Fig.~\ref{fig:diagrams}.  There are three components.  First in Fig.~\ref{fig:diagrams}(a) we have the perturbative \emph{resolved} Pomeron contribution in which the virtualities of the $t$-channel partons are strongly ordered as required by DGLAP evolution: $\mu_0^2\ll \ldots \ll \mu^2\ll \ldots \ll Q^2$.  The scale $\mu^2$ at which the Pomeron-to-parton splitting occurs can vary between $\mu_0^2\sim 1$ GeV$^2$ and the factorisation scale $Q^2$.  For $\mu^2<\mu_0^2$, the representation of the Pomeron as a perturbative parton ladder is no longer valid and instead, in the lack of a precise theory of non-perturbative QCD, we appeal to Regge theory where the soft Pomeron is a Regge pole with intercept $\alpha_\Pom(0)\simeq 1.08$ \cite{Donnachie:1992ny}; see Fig.~\ref{fig:diagrams}(b).  In addition to the \emph{resolved} Pomeron contributions of Figs.~\ref{fig:diagrams}(a,b), we must also account for the \emph{direct} interaction of the perturbative Pomeron in the hard subprocess, Fig.~\ref{fig:diagrams}(c), where there is no DGLAP evolution in the upper part of the diagram.

Each diffractive DIS event is described by four variables, usually taken as $x_\Pom$, $\beta$, $Q^2$ and $t$.  Here, $x_\Pom=1-{p^\prime}^+/p^+$ is the fraction of the proton's light-cone momentum transferred through the rapidity gap, $\beta=\xB/x_\Pom$ is the fraction of the Pomeron's light-cone momentum carried by the struck quark (at LO), and $\xB = Q^2/(2p\cdot q)$ is the Bjorken-$x$ variable. The data are integrated over the squared momentum transfer $t=(p-p^\prime)^2$. The variables are shown in Fig.~\ref{fig:diagrams}(a). In the pQCD approach, the diffractive structure function $F_2^{{\rm D}(3)}$ is given by \cite{Martin:2005hd,Watt:2005gt}
\begin{equation}
  F_{2}^{{\rm D}(3)}(x_\Pom,\beta,Q^2) = \sum_{a=q,g} C_{2,a}\otimes a^{{\rm D}} ~+~ \sum_{\Pom=G,S,GS}C_{2,\Pom}, \label{eq:f2d3}
\end{equation}
where the first \emph{resolved} Pomeron term corresponds to Figs.~\ref{fig:diagrams}(a,b) while the second \emph{direct} Pomeron term corresponds to Fig.~\ref{fig:diagrams}(c).  The $C_{2,a}$ are the same coefficient functions as in inclusive DIS.  The diffractive\footnote{To be precise, events with a large rapidity gap are described not only by Pomeron exchange, but also exchange of secondary Reggeons.  From this point of view, LRG events are not strictly diffractive.  However, the parton densities which originate from secondary Reggeons satisfy the same inhomogeneous evolution equations \eqref{eq:evol}.  Moreover, even in the inhomogeneous term of \eqref{eq:evol} the Pomeron flux $f_\Pom(x_\Pom;\mu^2)$, written in \eqref{eq:fg} in terms of the inclusive parton densities, also contains some admixture from secondary Reggeons and not just from the rightmost (Pomeron) singularity in the complex angular momentum plane.  So everywhere the word ``diffractive'' and the index $\Pom$ should be regarded as a symbolic, rather than a literal, notation.} parton distribution functions (DPDFs), $a^{\rm D}=z q^{\rm D}$ or $z g^{\rm D}$, are obtained at a scale $Q^2$ by evolving up from $\mu_0^2$ with the inhomogeneous evolution equations \cite{Martin:2005hd,Watt:2005gt},
\begin{equation} \label{eq:evol}
    \frac{\partial a^{\rm D}(x_\Pom,z,\mu^2)}{\partial \ln \mu^2}
    = \sum_{a^\prime=q,g}P_{aa^\prime}\otimes a^{\prime\,{\rm D}}~+~\sum_{\Pom=G,S,GS} P_{a\Pom}(z)\,f_\Pom(x_\Pom;\mu^2).
  \end{equation}
Here, $z\in[\beta,1]$ is the fraction of the Pomeron's light-cone momentum.  The first term on the right-hand-side of \eqref{eq:evol} generates the upper part of the ladder in Fig.~\ref{fig:diagrams}(a), that is, above $\mu^2$, via the usual (homogeneous) DGLAP evolution with parton-to-parton splitting functions, $P_{aa^\prime}$.  The second (inhomogeneous) term describes the cell where the first $t$-channel single parton is generated directly by the Pomeron-to-parton splitting functions, $P_{a\Pom}$, at a scale $\mu^2$.  In other words, these functions describe the transition from the two $t$-channel partons (that is, the Pomeron) to a single $t$-channel parton.  The LO expressions for $P_{a\Pom}$ are given in Ref.~\cite{Martin:2005hd}.  The lower parton ladder in Fig.~\ref{fig:diagrams}(a) plays the r\^ole of a perturbative Pomeron flux factor, $f_\Pom(x_\Pom;\mu^2)$.  These flux factors may be calculated from the familiar global parton densities \cite{Martin:2005hd}.  For example, if the two $t$-channel partons are gluons, then
\begin{equation}
  f_{\Pom=G}(x_\Pom;\mu^2)~=~\frac{1}{x_\Pom B_D}\left[R_g\frac{\alpha_S(\mu^2)}{\mu}x_\Pom g(x_\Pom,\mu^2)\right]^2,
  \label{eq:fg}
\end{equation}
where the diffractive slope parameter $B_D$ comes from the $t$ integration, and $R_g$ from the skewedness of the gluon distribution \cite{Shuvaev:1999ce}.  The notation $\Pom=G,S,GS$ denotes whether these two $t$-channel partons are gluons ($\Pom=G$) or sea-quarks ($\Pom=S$), while the interference term is denoted by $\Pom=GS$.  We neglect the case where the two $t$-channel partons are valence quarks, assuming that $x_\Pom\ll 1$.  We will take $B_D=6$ GeV$^{-2}$ \cite{H1FPS}, apart from for heavy quark production via a \emph{direct} Pomeron mechanism, Fig.~\ref{fig:diagrams}(c), where we take $B_D=4$ GeV$^{-2}$ \cite{Chekanov:2002xi,Aktas:2005xu}.

Now we come to an important observation. If we neglect the $\mu^2$ dependence of $g(x_\Pom,\mu^2)$ in \eqref{eq:fg}, then $f_{\Pom=G}(x_\Pom;\mu^2) \sim 1/\mu^2$ and the inhomogeneous term would be a higher-twist power-like correction, which is usually neglected and so collinear factorisation would hold in DDIS. However, as $x_\Pom \to 0$, the gluon density grows as $(\mu^2)^\gamma$, with anomalous dimension $\gamma=0.5$, while in the HERA domain $\gamma \sim 0.3$. This compensates for the $1/\mu^2$ power-like suppression in \eqref{eq:fg}. We therefore cannot neglect the inhomogeneous term in \eqref{eq:evol}.

Using only the perturbative framework, that is, just the diagrams in Figs.~\ref{fig:diagrams}(a,c), down to a very low scale would, of course, be unreliable. However, it is reasonable to assume that all the contributions originating from these low scales, $\mu^2<\mu_0^2$, are parameterised in terms of an unknown non-perturbative input, which is schematically presented in Fig.~\ref{fig:diagrams}(b). Unlike the unknown input required to analyse inclusive DIS which depends on a single momentum fraction, here we have an unknown input which depends on two momentum fractions, $x_\Pom$ and $z$. In this non-perturbative domain it is reasonable to assume a Regge behaviour, in which the $x_\Pom$ dependence is given by the sum of Regge poles (including possible interference terms), and that each pole generates input diffractive parton distributions with a freely-parameterised $z$ behaviour, where the parameters are to be determined by fitting to the DDIS data. Note that for very low $x_\Pom$ it would be sufficient to keep just Pomeron exchange.

Thus in the pQCD approach we consider the DPDFs to be a sum of Pomeron and secondary Reggeon contributions,
\begin{equation}
  a^{\rm D}(x_\Pom,z,Q^2) = a^{{\rm D},\Pom}(x_\Pom,z,Q^2) + a^{{\rm D},\Reg}(x_\Pom,z,Q^2),
\end{equation}
where we have neglected the possible interference term, as is the usual practice \cite{H1LRG}.  We parameterise the input quark-singlet and gluon DPDFs for the leading Pomeron pole at a starting scale $\mu_0^2$ in the form:
\begin{align}
  z\Sigma^{{\rm D},\Pom}(x_\Pom,z,\mu_0^2) &= f_\Pom(x_\Pom)\; A_q\,z^{B_q}(1-z)^{C_q},\label{eq:inputq} \\
  z g^{{\rm D},\Pom}(x_\Pom,z,\mu_0^2) &= f_\Pom(x_\Pom)\; A_g\,z^{B_g}(1-z)^{C_g},\label{eq:inputg}
\end{align}
where the Pomeron flux factor is taken from Regge phenomenology \cite{H1LRG},
\begin{equation} \label{eq:pomflux}
  f_\Pom(x_\Pom) = A_\Pom\,\int_{t_\mathrm{cut}}^{t_\mathrm{min}}\!\dif{t}\quad\mathrm{e}^{B_\Pom\,t}\;x_\Pom^{1-2\alpha_\Pom(t)},
\end{equation}
with $\alpha_\Pom(t) = \alpha_\Pom(0) + \alpha_\Pom^\prime\,t$.  The secondary Reggeon contributions may be included in a similar way.  However, in practice, the DDIS data at large $x_\Pom$ are insufficient to constrain the input secondary Reggeon parameters.  Instead, following H1 \cite{H1LRG}, we fix the $z$ and $Q^2$ dependence of $a^{{\rm D},\Reg}(x_\Pom,z,Q^2)$ to be the same as the pionic parton distributions, that is, we neglect the inhomogeneous terms in the evolution, and fit only an overall normalisation factor $c_\Reg$.  This is reasonable since the secondary Reggeons are important only for larger $x_\Pom$, where the anomalous dimension is not large enough to cancel the power suppression in the corresponding perturbative Pomeron flux factors.  Moreover, in the perturbative domain, the secondary Reggeons are mainly non-singlet quarks which, even at very small $x_\Pom$, have a small (or negative) anomalous dimension.  This completes the summary of the pQCD procedure which is necessary to describe DDIS.

Now we introduce the Regge factorisation approach and relate it to the above pQCD procedure. In the ``Regge'' description the last term in \eqref{eq:f2d3} and the inhomogeneous term in \eqref{eq:evol} are neglected.  The factorising Regge input, \eqref{eq:inputq} and \eqref{eq:inputg}, is used in the remaining, now homogeneous, DGLAP evolution equations, \eqref{eq:evol}.  This assumption is justified for sufficiently large $\mu_0^2$ and fixed $x_\Pom$. The smaller the value of $x_\Pom$, the larger the value of $\mu_0^2$ has to be to justify the pure Regge approach\footnote{By ``pure Regge approach'' we mean that we can neglect the inhomogeneous term in \eqref{eq:evol} and describe the DPDFs $a^{\rm D}$ by the linear evolution equations starting from input generated by some set of Regge poles. Of course, at a large scale it may be necessary to have a more complicated structure in the complex angular momentum plane than simply the Pomeron and secondary Reggeon poles.}.  What happens if the Regge approach is applied for smaller values of $\mu_0^2$?  The non-negligible last term in \eqref{eq:evol} is attempting to be mimicked by enlarging the Regge input at $\mu_0^2$.  As a consequence we obtain the wrong $\mu^2$ dependence of the DPDFs.  We can readily glimpse what the result will be.  Experimentally we measure $F_{2}^{{\rm D}(3)}$, which is dominantly determined by the diffractive quark density.  Thus, to mimic the $\mu^2$ dependence of the last term in \eqref{eq:evol} in a pure Regge fit, we would expect to obtain a larger gluon at relatively large $z$ so as to provide a faster growth of the quark density with increasing $\mu^2$.

How different are the diffractive parton distributions obtained from fits to the DDIS data using the ``pQCD'' and the simpler ``Regge'' approaches?  To see this, we perform fits to the latest H1 LRG data \cite{H1LRG} for the diffractive reduced cross section $\sigma_r^{{\rm D}(3)}$.  We choose this particular data set because H1 have already performed a ``Regge'' parton analysis to precisely these data with $M_X>2$ GeV and $Q^2\ge Q_{\rm min}^2 = 8.5$ GeV$^2$.  We will use the same kinematic cuts, in order to allow a direct comparison with the H1 2006 analysis \cite{H1LRG}.  We take $\alpha_\Pom(0)$ as a free parameter in \eqref{eq:pomflux} and fix $B_\Pom=5.5$ GeV$^{-2}$ and $\alpha_\Pom^\prime=0.06$ GeV$^{-2}$ \cite{H1FPS}, as in the H1 fit.  Moreover, the secondary Reggeon contributions are also included in a similar way to the H1 2006 analysis \cite{H1LRG}, with $\alpha_\Reg(0)=0.5$, $\alpha_\Reg^\prime=0.3$ GeV$^{-2}$ and $B_\Reg=1.6$ GeV$^{-2}$ \cite{H1FPS}, but using the GRV NLO pionic parton distributions \cite{Gluck:1991ey} rather than the Owens LO set \cite{Owens:1984zj}.  The number of parameters in the pQCD and Regge descriptions are the same, since the inhomogeneous term and \emph{direct} Pomeron contributions are calculated explicitly \cite{Martin:2005hd}, at leading order, and have no free parameters.  We choose to take the input parameterisations at $\mu_0^2=2$ GeV$^2$, again to keep close to the value used in the H1 analysis \cite{H1LRG}, and because the inclusive parton densities which appear in the pQCD approach are not well known below $2$ GeV$^2$ where there are no data included in the global fits.  We use the MRST2004F3 NLO PDFs \cite{Martin:2006qz} in the pQCD fits.  The inhomogeneous evolution is implemented via a modified version of the \textsc{qcdnum} \cite{QCDNUM} NLO DGLAP program.

For the Pomeron contribution, we include heavy quarks in the fixed flavour number scheme (FFNS), where there are no heavy quark DPDFs, with $m_c = 1.43$ GeV and $m_b = 4.3$ GeV, and use a three-flavour $\alpha_S$ with $\Lambda_{\rm QCD}^{(n_f=3)} = 407$ MeV \cite{Martin:2006qz}.  For the \emph{resolved} Pomeron case, heavy quarks are generated via photon--gluon fusion at NLO \cite{QCDNUM} with the renormalisation and factorisation scales taken to be $(Q^2+4m_h^2)$ where $h=c,b$.  For the secondary Reggeon case, the contribution to the diffractive structure functions from heavy quarks is included using the zero-mass variable flavour number scheme, and we take the same values of $\alpha_S$, $m_c$ and $m_b$ as in the determination of the pion PDFs \cite{Gluck:1991ey}.

The \emph{direct} Pomeron terms $C_{2,\Pom}=C_{T,\Pom} + C_{L,\Pom}$, that is, the second terms on the right-hand-side of \eqref{eq:f2d3}, are calculated at LO using:
\begin{gather}
  C_{T,\Pom=G} = \text{(A.27)} - \text{(A.28)},\qquad C_{L,\Pom=G} = \text{(A.36)},\\
  C_{T,\Pom=S} = \text{(A.68)} - \text{(A.69)},\qquad C_{L,\Pom=S} = \text{(A.74)},\\
  C_{T,\Pom=GS} = \text{(A.86)} - \text{(A.87)},\qquad C_{L,\Pom=GS} = \text{(A.91)}.
\end{gather}
where the equation numbers refer to the Appendix of Ref.~\cite{Martin:2005hd}.  For light quarks, the lower limit of the $\mu^2$ integration in these equations is replaced by $\mu_0^2$, since the contribution from $\mu^2<\mu_0^2$ is included in the input DPDFs taken at a scale $\mu_0^2$.  For $C_{T,\Pom}$ the light-quark contributions in the limit $\mu^2\ll Q^2$ are subtracted since they are already included in the first term of \eqref{eq:f2d3}.  This subtraction defines a choice of factorisation scheme.  There is no such subtraction for $C_{L,\Pom}$, which are purely higher-twist, or for the heavy quark contributions since we are working in the FFNS.

\begin{table}
  \centering
  \begin{tabular}{c|c|c}
    \hline\hline
    & ``Regge'' fit & ``pQCD'' fit \\ \hline
    $\chi^2/\mathrm{d.o.f.}$ & $142/(190-8)=0.78$ & $153/(190-8)=0.84$ \\ \hline
    $\alpha_\Pom(0)$ & $1.117\pm0.009$ & $1.106\pm0.005$ \\
    $c_\Reg$ & $(1.21\pm0.34)\times 10^{-3}$ & $(1.09\pm0.20)\times 10^{-3}$ \\
    $A_q$ & $0.86\pm0.26$ & $0.57\pm0.29$   \\
    $B_q$ & $2.21\pm0.35$ & $1.60\pm0.05$   \\
    $C_q$ & $0.63\pm0.16$ & $0.65\pm0.04$   \\
    $A_g$ & $0.14\pm0.08$ & $0.22\pm0.02$   \\
    $B_g$ & $0.02\pm0.22$ & $0.31\pm0.06$   \\
    $C_g$ & $-0.92\pm0.42$ & $-0.19\pm0.04$ \\
    \hline\hline
  \end{tabular}
  \caption{Input parameters (corresponding to $M_Y=m_p$) determined in the Regge and pQCD fits to H1 LRG data \cite{H1LRG} with $Q_{\rm min}^2 = 8.5$ GeV$^2$.  Notice that the $\Delta\chi^2=1$ errors on the parameters are generally smaller in the pQCD fit.}
  \label{tab:params}
\end{table}

\begin{figure}
  \begin{center}
    \includegraphics[width=\textwidth]{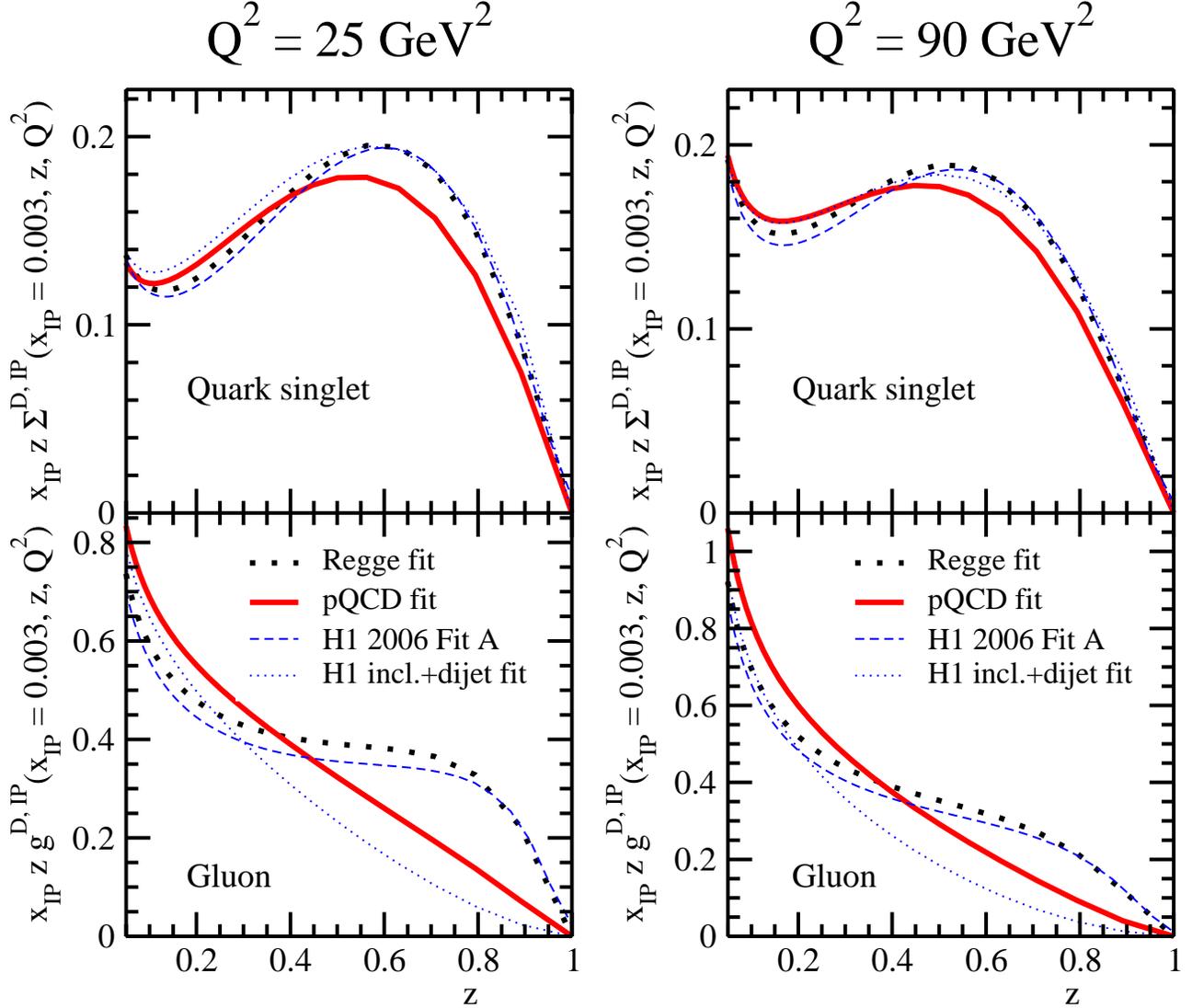}
    \caption{The ``Regge'' and ``pQCD'' DPDFs obtained from fitting to the H1 LRG data \cite{H1LRG} with $Q_{\rm min}^2=8.5$ GeV$^2$ compared to the H1 2006 Fit A \cite{H1LRG}.  All the DPDFs plotted here are normalised to $M_Y < 1.6$ GeV by multiplying by a factor 1.23 relative to $M_Y = m_p$.  Only the Pomeron contribution is shown and not the secondary Reggeon contributions which are negligible at the value of $x_\Pom=0.003$ chosen here.  Also shown are the DPDFs obtained by H1 \cite{H1jets}, using the Regge factorisation approach, from a preliminary combined analysis of their inclusive DDIS data \cite{H1LRG} with preliminary data on diffractive dijet production \cite{H1jets}.}
    \label{fig:dpdf}
  \end{center}
\end{figure}

Examples of the diffractive parton densities obtained from these fits to the H1 LRG data are shown in Fig.~\ref{fig:dpdf}.  During the fits, statistical and systematic experimental errors were added in quadrature.  An overall factor of 1.23 \cite{H1FPS} was applied to the theoretical prediction to account for the effect of proton dissociation up to masses $M_Y<1.6$ GeV present in the H1 LRG data \cite{H1LRG}.  The parameters are given in Table \ref{tab:params} and correspond to $M_Y=m_p$.  As is to be expected, there are only very small differences between our ``Regge'' fit and the H1 2006 Fit A.\footnote{These are due to minor details in the two fits such as, for example, the treatment of experimental errors, the precise form of the input parameterisation, the choices for $\alpha_S$, $m_c$, $m_b$, the heavy quark renormalisation and factorisation scales, and the pion PDFs.  However, as seen in Fig.~\ref{fig:dpdf}, these minor details are not important.}  The value of the $\alpha_\Pom(0)$ parameter in the pQCD fit is $1.106\pm0.005$.  It is a bit higher than the value of the intercept 1.08 \cite{Donnachie:1992ny} coming from soft hadron phenomenology, since the input distributions are taken at $\mu_0^2=2$ GeV$^2$, and already account for some pQCD contribution which is known to be present at this scale.  For example, in going from $\rho$ to $J/\psi$ photoproduction, the effective Pomeron intercept increases from about 1.09 \cite{Breitweg:1999jy,H1rho} to 1.20 \cite{Chekanov:2002xi,Aktas:2005xu}.  Note that a detailed comparison of the intercepts would require a study of the correlation between the fitted $\alpha_\Pom(0)$ and the assumed $\alpha_\Pom^\prime$, whereas here we have fixed $\alpha_\Pom^\prime = 0.06$ GeV$^{-2}$ to agree with the H1 analysis \cite{H1LRG}.

From Fig.~\ref{fig:dpdf} we see that the diffractive quark singlet distribution, $\Sigma^{\rm D,\Pom}$, is relatively stable in going from the pQCD approach to the simplified Regge approach.  However, as anticipated, the diffractive gluon distribution, $g^{\rm D,\Pom}$, is greatly suppressed for $z\gtrsim 0.5$ in the pQCD approach relative to the gluon obtained using the Regge approach.

The results obtained here are stable to the inclusion of DDIS data from the H1 FPS \cite{H1FPS}, the ZEUS leading proton spectrometer (LPS) \cite{Chekanov:2004hy}, and the ZEUS diffractive charm reduced cross section \cite{Chekanov:2003gt}, as well as being stable to including H1 LRG data \cite{H1LRG} down to $Q^2 = 6.5$ GeV$^2$.  At an input scale of $\mu_0^2=2$ GeV$^2$, the results are also stable to an extra parameter in the input distributions in the form of an additional factor $(1+D_a\sqrt{z})$ and to the use of an alternative input parameterisation in terms of Chebyshev polynomials, although this is no longer the case if a higher input scale is used such as $\mu_0^2=6$ GeV$^2$.

When the forthcoming ZEUS LPS and LRG data \cite{ZEUSLRG} are released we will perform an analysis of the combined data sets, including also the measurements of diffractive dijet production.  The pQCD diffractive parton densities, $a^{{\rm D},\Pom}$ and $a^{{\rm D},\Reg}$, of the present analysis are publically available\footnote{\texttt{http://durpdg.dur.ac.uk/hepdata/mrw.html}} in the form of three-dimensional grids in $(x_\Pom,z,Q^2)$, together with the corresponding predictions for the diffractive structure functions $F_2^{{\rm D}(3)}$, $F_L^{{\rm D}(3)}$, $F_2^{{\rm D}(3),c\bar{c}}$ and $F_L^{{\rm D}(3),c\bar{c}}$.  The \emph{resolved} Pomeron, \emph{direct} Pomeron and secondary Reggeon contributions to the diffractive structure functions are provided separately in each of these four cases.  Measurements of the diffractive charm reduced cross section and the fractional contribution of charm to the total diffractive cross section have recently been shown to be in excellent agreement with these predictions; see Figs.~12 and 13 of Ref.~\cite{H1charm}.

The H1 Collaboration have recently presented a preliminary combined fit \cite{H1jets}, within the Regge factorisation framework, of their inclusive DDIS data \cite{H1LRG} with preliminary data on diffractive dijet production \cite{H1jets}.  The DPDFs from this combined fit, shown in Fig.~\ref{fig:dpdf}, have a much smaller diffractive gluon density at moderate to high $z$ compared to the H1 2006 Fit A.  Moreover, the $\chi^2$ for the 190 inclusive DDIS points increases from 158 (H1 2006 Fit A) to 169 (H1 combined fit) on inclusion of the dijet data.  Therefore, the gluon density determined indirectly from the inclusive DDIS data, under the assumption of pure DGLAP evolution, is different from the gluon density preferred by the dijet data.  On the other hand, our pQCD fit naturally has a smaller gluon density at moderate to high $z$ than the H1 2006 Fit A, due to the presence of the extra inhomogeneous term in the evolution equations.  Therefore, the apparent tension between the inclusive DDIS and dijet data in the Regge factorisation approach is partly alleviated by the inclusion of the perturbative Pomeron terms.  As a consequence, we expect that our pQCD fit will give a better description of the diffractive dijet data than the H1 2006 Fit A.

\section*{Acknowledgements}
We thank Paul Newman for valuable discussions.  MGR and GW would like to thank the IPPP at the University of Durham for hospitality.  This work was supported by the Royal Society, by INTAS grant 05-103-7515, by grant RFBR 04-02-16073 and by the Federal Program of the Russian Ministry of Industry, Science and Technology SS-1124.2003.2.

\end{document}